\documentclass[twocolumn,floatfix, aps,prd]{revtex4-2}
\usepackage{graphicx}
\usepackage{dcolumn}
\usepackage{amsmath}
\usepackage{epsfig}
\usepackage{subfigure}
\usepackage{longtable}
\usepackage{bm}
\usepackage{relsize}
\usepackage{multirow}
\usepackage{lineno}
\usepackage{xcolor}
\usepackage{hyperref}
\usepackage{cleveref}
\graphicspath{{./plots/}} 

\begin{document}
    \title{Recent spectroscopy studies at Belle \footnote{Presented at CHARM2023, Siegen, Germany}}
    \author{Dmytro Meleshko}
    \author{Elisabetta Prencipe}
    \author{Jens Sören Lange}
    \affiliation{Justus-Liebig-Universität Gießen, Gießen, Germany}

    \date{\today}
    
    \begin{abstract}
        The analyses of states with double $cs$ content and the search for exotics have recently gained much attention. The Belle experiment collected roughly 1 ab$^{-1}$ integrated luminosity data. While Belle II data-taking is in progress, we have performed a new search for exotic states and cross-section measurements with the full Belle data sets. Here we review the recent analysis of: (a) $e^+e^-\to D_s^{(*)+}+D_{sJ}^-$ + $c.c$. from both $\Upsilon(2S)$ decays and continuum production at 10.52 GeV, using the Belle detector at KEKB; (b) the analysis of $e^+e^- \to \eta_c J/\psi$ + $c.c.$ and search for double charmonium states; (c) the study of $e^+e^- \to D_s^+D_{s0}(2317)^-$ + $c.c.$ and $e^+e^- \to D_s^+D_{s1}(2460)^-$ + $c.c.$ + anything else, in the continuum. Born cross-sections are evaluated, and a possible confirmation of the states seen in the invariant mass system of $J/\psi\phi$ by LHCb in $B$ decays has been investigated.
    \end{abstract}

    \maketitle
    
    \section{Introduction}
    
    B-factories have discovered or confirmed many of the resonances currently observed in the charmonium-like spectrum. These states, presently referred to as “XYZ” states, could be observed thanks to the clean environment offered by electron-positron experiments and large $e^+e^-\to q\bar{q}$ production rates (about 40\% of the total hadronic production at $\sqrt{s}=10.52$ GeV). However, many candidates observed above the $D\bar{D}/B\bar{B}$ threshold do not match current QCD predictions. Hence, substantial studies of already discovered states and the discovery of new particles are essential for better theoretical interpretation of the quarkonium(-like) spectrum. 

    Several exotic states have been discovered in the decays into two heavy-flavour mesons and/or a quarkonium and one or two light hadrons \cite{Brambilla_2020}.
    
    Charmonium-like states with $J^{PC}=1^{--}$ named now conventionally as $\psi(4260)$ \cite{Y4260_BaBar}, $\psi(4360)$ \cite{Y4360_BaBar, Y4360_Belle} and $\psi(4660)$ \cite{Y4360_Belle} have been addressed for long time as $Y$ states. For example, the $\psi(4260)$ was originally discovered by BaBar in the initial state radiation (ISR) process $e^+e^-\to\gamma_{ISR} J/\psi \pi^+ \pi^-$ and was reported to have a mass of $4259\pm8^{+2}_{-6}$ MeV/c$^2$. It was later confirmed by CLEO \cite{Y4260_CLEO} and Belle \cite{Y4260_Belle} in the same process. The lattice quantum chromodynamics (QCD) calculation in Ref. \cite{latticeQCD_Y} predicted the mass of $\psi(4260)$ to be 4238$\pm$31 MeV/c$^2$ under the hypothesis of being a molecule. Moreover, this model derived another two $cs\bar{c}\bar{s}$ and $cc\bar{c}\bar{c}$ states with masses of 4450$\pm$100 MV/c$^2$ and 6400$\pm$50 MeV/c$^2$, respectively. A dedicated BESIII analysis has revealed that the observed resonance in the vicinity of 4260 MeV/c$^2$ is not a single state but two close-by peaks are likely observed \cite{Y4230_BSIII}. The lower-mass resonance was reported to have a narrower width and significance of more than 7.6$\sigma$; it is known today as $\psi(4230)$. This state was later observed in $e^+e^-\to\pi^+\pi^-h_c$ \cite{Y4230_BESIII_confirm_1,Y4230_BESIII_confirm_2}, $e^+e^-\to\omega\chi_{c0}$ \cite{Y4230_BESIII_chicomega} and $\pi\bar{D}D^*+c.c.$ decays \cite{Y4230_BESIII_pibarDDst}, as well as in $\eta$ transition to lower charmonium states \cite{Y4230_BESIII_etaJpsi,Y4230_BESIII_etaprimJpsi}.

    A recent study of $e^+e^-\to K^+K^-J/\psi$ cross-section reported observation of a narrow resonance with $M=4487.8\pm13.3\pm24.1$ MeV/c$^2$ \cite{BESIII_KKJpsi}, which is very close to the aforementioned lattice QCD prediction for the $cs\bar{c}\bar{s}$ state. Another cross-section measurement in Belle via ISR reported the observation of a resonance with a mass around 4620 MeV/c$^2$ in $e^+e^-\to D_s^+D_{s1}(2536)^-+c.c.$ \cite{SenJia_1} and $e^+e^-\to D_s^+D_{s2}^*(2573)^-+c.c.$ \cite{SenJia_2} processes. Meanwhile, recent studies of $B^+\to K^+\phi J/\psi$ decays in LHCb showed the observation of several resonances in the $\phi J/\psi$ system \cite{LHCb_jpsiPhiK_1,LHCb_jpsiPhiK_2}, which can be interpreted as $cs\bar{c}\bar{s}$ states. Furthermore, the observed narrow structure in the double-$J/\psi$ system with a mass around 6900 MeV/c$^2$, dubbed $X(6900)$, is considered to be the most viable candidate for the remaining $cc\bar{c}\bar{c}$ state predicted by the same lattice QCD calculation. Observation of the $X(6900)$ was later confirmed by ATLAS and CMS \cite{CMS_X6900}. It is though thought that the interaction between two charmonia might be not strong enough to form a tight bound state \cite{doubl_charmonium_1}, so the compact tetraquark model, $e.g.$ compact diquark anti-diquark $[QQ][\bar{Q}\bar{Q}]$ structure, is adopted by many theoretical studies \cite{doubl_charmonium_2}.
    
    A study of the open-charm production in hadronic decays of bottomonium, including Okubo-Zweig-Iizuka suppressed hadronic decays of $\Upsilon(\text{nS})$, can be an effective test for the QCD prediction on the production of heavy-quarkonia. Even though available measurements of charm hadrons in $\Upsilon(\text{nS})$ are poor yet, some analyses $i.e.$ the BaBar cross-section measurement of the $\Upsilon(1S)\to D^{*+}X$ process \cite{BaBar_DstX_crosssect}, highlight the necessity of further studies to accommodate contributions of higher orders \cite{DstX_theoryRef}.

    The search for exotic states in the continuum enhances the possibility to complement conventional studies of $B$-decays and ISR processes. The advantage of a search in the continuum is that all quantum numbers are directly accessible because the invariant mass of a system is analyzed in the recoil of anything else. In this case, all the available data samples at the various center of mass (c.m.) energies of $\Upsilon(1,2,3,4,5S)$ can be used. For example, the Belle experiment can potentially access possible molecular states with $J^P$ = $0^-$ and $2^-$ bound by $\eta$ exchange, 
    which were predicted by Ref. \cite{eta_exchange} by performing studies in the continuum, while they can not be observed by LHCb in the $J/\psi \phi$ invariant mass, possibly due to the quantum numbers ($J/\psi$ and $\phi$ are vectors).

    \begin{table*}[!hbt]
        \begin{tabular}{ccccccc}
            \hline 
            \hline 
            & $\Upsilon(1S)$ & $\Upsilon(2S)$ & $\Upsilon(3S)$ & $10.52 \mathrm{GeV}$ & $\Upsilon(4S)$ & $\Upsilon(5S)$ \\
            \hline 
            $\mathcal{L}\left[\mathrm{fb}^{-1}\right]$ & 5.7 & 24.9 & 2.9 & 89.4 & 711.0 & 121.4 \\
            $N^{\text{exc}}$ & $0.7_{-0.9}^{+1.5}$ & $6.2_{-2.3}^{+3.1}$ & $<1.9$ & $2.6_{-2.5}^{+3.5}$ & $45.0_{-8.2}^{+8.9}$ & $6.5_{-2.7}^{+3.4}$ \\
            $\epsilon^{\text {exc }}$ & $8.3 \%$ & $6.9 \%$ & $5.7 \%$ & $5.6 \%$ & $5.6 \%$ & $5.4 \%$ \\
            $\sigma^{\text{exc}}[\mathrm{fb}]$ & $57_{-73}^{+122} \pm 6$ & $140_{-52}^{+70} \pm 14$ & $<442$ & $20_{-19}^{+27} \pm 6$ & $44_{-8}^{+9} \pm 5$ & $39_{-14}^{+20} \pm 7$ \\
            \hline
            $N^{\text{inc}}$ & 23.7$\pm$12.3 & 62.0$\pm$17.9 & 8.5$\pm$5.2 & 94.7$\pm$23.8 & 1116.2$\pm$62.9 & 91.1$\pm$21.5 \\
            $\epsilon^{\text{inc}}$ & $38.6 \%$ & $29.6 \%$ & $26.4 \%$ & $26.1 \%$ & $25.4 \%$ & $24.7 \%$ \\
            \hspace{-3mm} $\sigma^{\text{inc}}[\mathrm{fb}]$ & \hspace{-4mm} $89.1^{\pm46.2}_{\pm20.5}$ & \hspace{-3mm} $70.1^{\pm20.2}_{\pm8.9}$ & \hspace{-3mm} $91.8^{\pm56.2}_{\pm52.3}$ & \hspace{-3mm} $33.8^{\pm8.5}_{\pm2.8}$ & \hspace{-3mm} $52.1^{\pm2.9}_{\pm5.0}$ & \hspace{-3mm} $25.4^{\pm6.0}_{\pm2.8}$ \hspace{-3mm} \\
            \hline
            $\sigma^{\text{comb}}[\mathrm{fb}]$ & $78.3_{-4.0}^{+47.5}$ & $80.2 \pm20.4$ & $87.0_{-59.0}^{+71.0}$ & $32.5 \pm 8.5$ & $50.2 \pm 5.0$ & $27.5 \pm 6.1$ \\
            \hline
        \end{tabular}
        \caption{Signal yields and the measured cross-sections for different channels at selected c.m. energy points. The first uncertainties in $\sigma^{\text{exc}}$ are statistical (top for $\sigma^{\text{inc}}$) and the second are systematic (bottom for $\sigma^{\text{inc}}$). The $\sigma^{\text {comb }}$ total uncertainties are given, including the statistical and systematic \cite{etacJpsi}.}
        \label{tab:etacJpsi}
    \end{table*}

    The Belle detector is a large-solid-angle magnetic spectrometer \cite{Belle_detctor} using a silicon vertex detector, a 50-layer central drift chamber, an array of aerogel threshold Cherenkov counters, a barrel-like arrangement of time of flight scintillation counters, and an electromagnetic calorimeter (ECL) composed of CsI(Tl) crystals located inside a superconducting solenoid coil that provides a 1.5 T magnetic field. An iron flux return outside the coil is instrumented to detect $K_L^0$ mesons and identify muons. The Belle detector is well suited for radiative analyses and all analyses involving low-momentum pions and low-energy photons. It is possible because of its excellent calorimeter performance, which allows it to detect photons with energies down to about 50 MeV/$c$.

    \section{Search for the double-charmonium state with $\mathbf{\eta_cJ/\psi}$ at Belle \cite{etacJpsi}}

    A recent Belle study \cite{etacJpsi} reports results of a search for a predicted $cc\bar{c}\bar{c}$ state in decays into $\eta_c J/\psi$ in ISR process. Since this is the lowest mass combination of charmonia to which a vector $cc\bar{c}\bar{c}$ could decay, this process may have a relatively large branching fraction. A search was performed using the full Belle data sample (980 fb$^{-1}$), which was collected at $\Upsilon(\text{nS})$ resonances and in the continuum. The analysis strategy included measurement of the cross-section in the vicinity of the $\Upsilon(\text{nS})$ energy points to validate the analysis method and to perform a solid check for the next-to-next-to-leading-order calculation in the NRQCD approach \cite{zhang2023twoloop}. The results of these measurements were then extrapolated to the near-threshold region to search for $Y_{cc}$ in the region of interest based on the continuum prediction.

    Two distinct reconstruction methods were implemented in this analysis. One is an exclusive reconstruction of $\eta_c J/\psi$, and the other is an inclusive reconstruction using $J/\psi$ or $J/\psi\gamma_{ISR}$ for $\Upsilon(\text{nS})$ on/off resonance or near-threshold data. A dedicated selection was optimized to look for $J/\psi$ candidates in $e^+e^-$ and $\mu^+\mu^-$ decays, while $\eta_c$ candidates were selected in $p\bar{p}$, $p\bar{p}\pi^0$, $K^0_SK^\pm\pi^\mp$, $K^+K^-\pi^0$, $2(K^+K^-)$ and $2(\pi^+\pi^-\pi^0)$ combinations for the exclusive approach.

    Exclusive reconstruction, as expected, derived results with the limited number of reconstructed candidates. Meanwhile, the results of inclusive approach held higher potential. Thus, the $J/\psi$ recoil-mass was defined as $M_{\text{recoil}}(J/\psi)=\sqrt{|p_{e^+e^-}-p_{J/\psi}|^2}/c$, where $p$ is the four-momentum. 
    %A $M_{\text{recoil}}(J/\psi)$ plot corresponding to each c.m. energy is shown in Fig.~\ref{fig:etacJpsi_incl}. 
    A mass-constraint fit on the reconstructed $J/\psi$ candidates was additionally applied to improve resolution. Distinctive enhancements representing $\eta_c$, $\chi_{c0}$ and $\eta_c(2S)$ are observed in $M_{\text{recoil}}(J/\psi)$ plots, matching earlier Belle measurements \cite{PhysRevLett.98.082001, PhysRevD.90.112008}. Unbinned extended maximum likelihood fits were performed to the $M(\eta_c)$ and the $M_{\text{recoil}}(J/\psi)$ distributions to extract signal yield. Cross-sections are then calculated as $\sigma=N_{sig}/\epsilon\mathcal{L}\mathcal{B}(J/\psi\to\ell^+\ell^-)\mathcal{B}(\eta_c\to \text{6 channels})$ for the exclusive approach, where $N_{sig}$ is the number of signal events, $\epsilon$ is the reconstruction efficiency, $\mathcal{L}$ is the integrated luminosity, and $\mathcal{B}$ includes the corresponding branching fractions. The same equation is used for cross-section calculation in the inclusive analysis, but omitting the $\eta_c$ branching fractions. The calculated cross-sections for both analyses are shown in Tab.~\ref{tab:etacJpsi}. The results of cross-section measurements acquired with both methods were then combined.

    Two possible mechanisms for the $\eta_cJ/\psi$ production are expected: continuum and through an intermediate $\Upsilon(\text{nS})$ state. The fractions for both productions for the respective c.m. energies are consistent with those for the $e^+e^-\to\mu^+\mu^-$ process, and thus were utilized. The cross-section of the $e^+e^-\to\eta_cJ/\psi$ in continuum was fitted with an empirical function. The obtained fit was then extrapolated to the near-threshold region to estimate the continuum contribution if any signal were to be found there.

\begin{figure*}[!hbt]
        \includegraphics[width=0.85\textwidth]{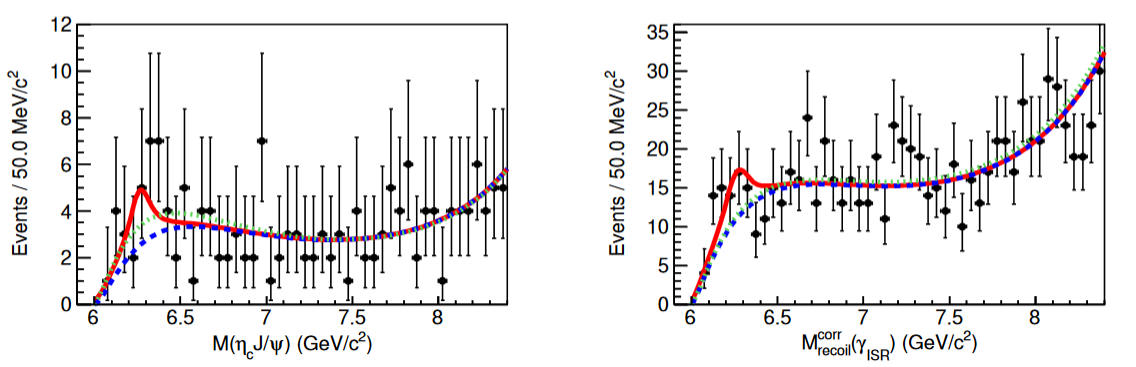}
        \caption{Simultaneous fit result to the invariant mass of $\eta_cJ/\psi$ (left) and the $\gamma$ recoil mass (right). In each panel, dots with error bars are from data, the red solid curve is the best fit result, the blue dashed curve the background component from the best fit, and the green dotted curve is the fit result without the signal components \cite{etacJpsi}.}
        \label{fig:sigma_etacJpsi_nt}
    \end{figure*}

    Following a similar analysis procedure, search for the $e^+e^-\to\eta_cJ/\psi$ events was performed in the near-threshold region. The events observed in inclusive and exclusive reconstructions are shown in Fig.~\ref{fig:sigma_etacJpsi_nt}. Common between two samples are removed from the inclusive reconstruction to avoid double counting. While a slight enhancement is observed at the threshold in the exclusive sample, no similar structure is seen in the recoil mass of $\gamma_{ISR}$. The significance of the Breit-Wigner peak component shown in Fig.~\ref{fig:sigma_etacJpsi_nt} is 2.1$\sigma$, with mass and width of (6267$\pm$43) MeV/c$^2$ and (121$\pm$72) MeV, respectively. The signal yields are 9$\pm$4 and 23$\pm$11 from the exclusive and inclusive methods, respectively.

    A simultaneous fit to the invariant mass of the reconstructed $\eta_c$ and the $\gamma_{ISR}J/\psi$ recoil mass has been performed for the samples with $M_{\text{recoil}}(\gamma)$ $\in$ [6.0, 6.4], [6.0, 6.5], and [6.0, 6.6] GeV/c$^2$. The significances of the $\eta_c$ components are 3.9, 3.3, and 3.5$\sigma$ for events from the three mass regions, respectively. No evident signals are found in those distributions, and the upper limits of the number of produced events in different $\eta_cJ/\psi$ mass regions are estimated at 90\% C.L.
    
\begin{figure}[!hbt]
        \centering
        \includegraphics[width=0.9\columnwidth]{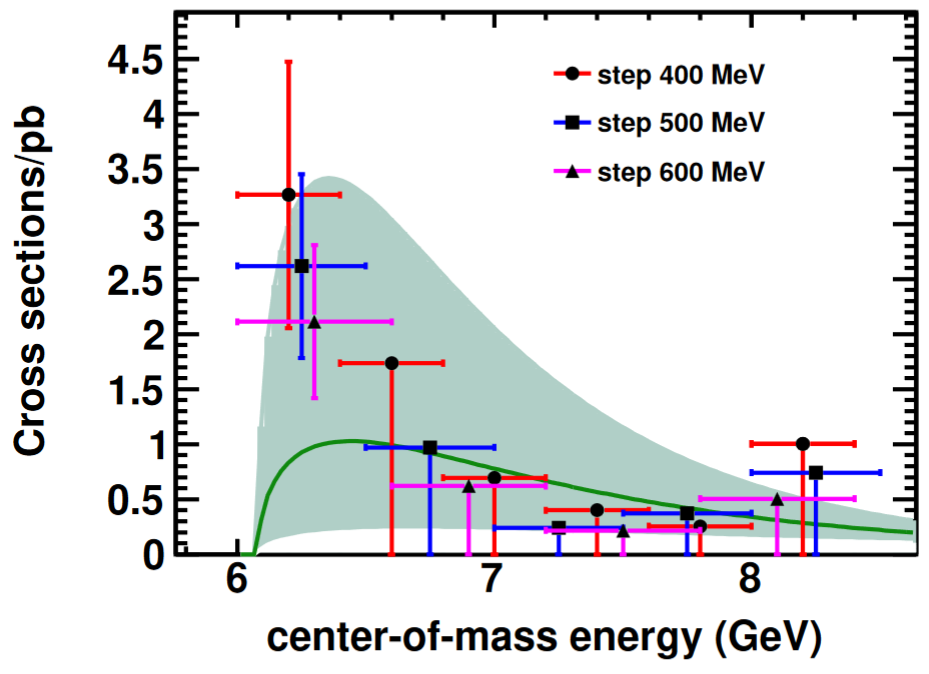}
        \caption{ Measured cross-sections of $e^+e^-\to\eta_cJ/\psi$ near the threshold. From left to right are the cross-sections measured in different step sizes (0.4, 0.5, 0.6 GeV/c$^2$) \cite{etacJpsi}.}
        \label{fig:etaJpsi_extrap}
    \end{figure}

    The effective luminosity in each mass region is calculated according to Ref. \cite{BENAYOUN_1999}. The cross-sections near $\eta_cJ/\psi$ mass threshold were estimated with an equation analogous to near-$\Upsilon(\text{nS})$ case with the corresponding errors (see Fig.~\ref{fig:etaJpsi_extrap}, in which the lineshape extrapolation of the measured cross-sections near $\Upsilon(\text{nS})$ resonances is depicted with a solid curve for comparison with the near-threshold measurements). Variation of the extrapolated parameters based on calculated uncertainties provides the $\pm\sigma$ area for the continuum prediction. The measured cross-sections near $\eta_cJ/\psi$ mass threshold were consistent with the extrapolations from the $\Upsilon(\text{nS})$ energy region according to their uncertainty.

    \section{Observation of charmed strange mesons pair production in $\mathbf{\Upsilon(2S)}$ decays and in $\mathbf{e^+e^-}$ annihilation at 10.52 GeV \cite{DsDsinY2S}}

    \begin{figure*}[!hbt]
        \centering
        \includegraphics[width=\textwidth]{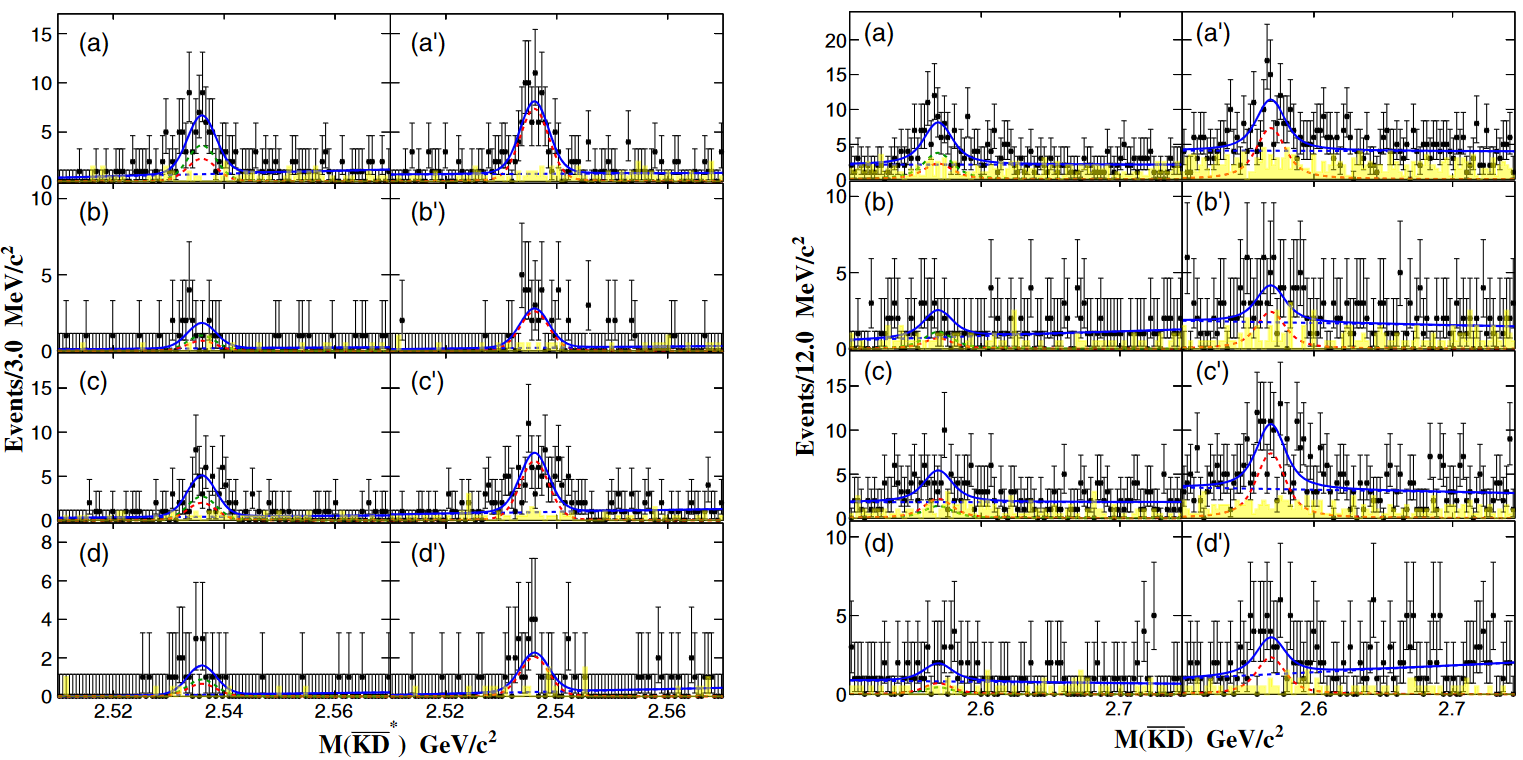}
        \caption{The invariant mass distributions of $\bar{K}\bar{D}^{(*)}$ calculated in the recoil mass for $D_s^{(*)+}$ in the: [left] (a) $D_s^{+} K^{-} \bar{D}^*(2007)^0$, (b) $D_s^{+} K_S^0 D^*(2010)^{-}, \quad$ (c) $D_s^{*+} K^{-} \bar{D}^*(2007)^0, \quad$ and (d) $D_s^{*+} K_S^0 D^*(2010)^{-}$; [right] (a) $D_s^{+} K^{-} \bar{D}^0$, (b) $D_s^{+} K_S^0 D^{-}$, (c) $D_s^{*+} K^{-} \bar{D}^0$, and (d) $D_s^{*+} K_S^0 D^{-}$ final states from the $\Upsilon(2S)$ data sample (a)-(d) and the continuum data sample at $10.52 \mathrm{GeV}\left(\mathrm{a}^{\prime}\right)-\left(\mathrm{d}^{\prime}\right)$. The shaded histograms show the backgrounds estimated from the normalized $D_s^{(*)+}$ mass sidebands. The solid curves show the best fit result; the dashed green ones are $D_{s1}(2536)^{-}$ signals in $\Upsilon(2S)$ decays, and the dashed red curves are the $D_{s1}(2536)^{-}$ signals in continuum production at 10.02 GeV (a$^\prime$)-(d$^\prime$) and 10.52 GeV \cite{DsDsinY2S}.}
        \label{fig:DsDs_KD}
    \end{figure*}

    Studies of charm hadrons production in $\Upsilon(\text{nS})$ decays are scarce. Meanwhile, the Belle experiment has accumulated 24.7 fb$^{-1}$ of data at the c.m. energy of $\Upsilon(2S)$ resonance and 89.7 fb$^{-1}$ of data in continuum at $\sqrt{s}=10.52$ GeV. Use of these data samples could potentially give an opportunity to separate the dynamics of electromagnetic and strong charmed hadron production at the on- and off-resonance energy. Therefore, search for $D_s^{(*)+}D_{sJ}^-$ with the subsequent $D_{sJ}^-$ decay into $\bar{K}\bar{D}^{(*)}$ has been recently performed in Belle. The following $D_{sJ}^-$ states were included in the analysis: $D_{s1}(2536)^-$ and $D_{s2}^*(2573)^-$. These states decay with the emission of a kaon, and thus can be analysed with the partial reconstruction method. The tagging $D_s^+$ was fully reconstructed in decays into $\phi\pi^+$, $K_S^0K^+$, $\bar{K}^*(892)^0K^+$, $\rho^+\phi$, $\eta\pi^+$ and $\eta^\prime\pi^+$; while only a decay into $D_s^+\gamma$ was considered for the tagging $D_s^*$. On the contrary, the recoiling $D_{sJ}^-$ candidates were selected by combining a reconstructed kaon and a $\bar{D}^{(*)}$ recoiling against the $D_s^{(*)+}$--$\bar{K}$ system. This circumvented the problem of low efficiencies for reconstructing $D$ mesons associated with the large variety of possible decay processes.

    After development and application of a designated selection for the exclusive sub-decays, inclusive $\bar{D}^{(*)}$ decay was studied in a recoil and a possible $D_{sJ}^-$ production was isolated in the $\bar{D}^{(*)}\bar{K}$ final states in the recoil against $D_s^{(*)+}$. Clear bands are observed in these distributions corresponding to the production of the $D_{s1}(2536)^-$ signal in the $\bar{D}^*(2007)^0K^-$ and the $D^*(2010)^-K^0_S$ final states, and the $D_{s2}^*(2573)^-$ signal production in $\bar{D}^0K^-$ and $D^-K^0_S$. The mass resolutions of $M_{\bar{D}^{(*)}\bar{K}}^{\text{recoil}}$ and  $M_{\bar{D}^{(*)}}^{\text{recoil}}$ distributions were large ($\approx$ 50 MeV/c$^2$) due to the shared variables used to determine recoil masses. The corrected mass was utilized instead to improve mass resolution: $M_{\bar{K}\bar{D}^{(*)}}=M_{D_s^{(*)+}}^{\text{recoil}}-M_{D_s^{(*)+}\bar{K}}^{\text{recoil}}+m_{\bar{D}^{(*)}}$. In this way, the uncertainties due to the 4-momentum of final states from $D_s^{(*)+}$ decays are significantly reduced. According to the MC studies, the resolution of the $\Delta M^{\text{recoil}}\equiv M_{\bar{D}^{(*)}\bar{K}}^{\text{recoil}} - M_{\bar{D}^{(*)}}^{\text{recoil}}$ distribution that can be reached in this way is less than to 5 MeV/c$^2$ for all $D_s^{(*)+}D_{sJ}^-$ final states. Fig.~\ref{fig:DsDs_KD} shows the distributions for $\Delta M^{\text{recoil}}+m_{\bar{D}^*}$ for $M_{\bar{K}\bar{D}^{(*)}}$ and $\Delta M^{\text{recoil}}+m_{\bar{D}}$ for $M_{\bar{K}\bar{D}}$ for the two data samples. Clear $D_{s1}(2536)^-$ and $D_{s2}^*(2573)^-$ signals are observed for both data samples.

    The $D_{sJ}$ signal yield in $\Upsilon(2S)$ decays and in continuum ($N_{\Upsilon(2S)}^{\text{sig}}$ and $N_{\text{cont}}^{\text{sig}}$) were estimated by fitting $M_{\bar{K}\bar{D}^{(*)}}$ distributions simultaneously with the common isospin ratios between the $K^0_SD^{(*)}$ and $K^-\bar{D}^{(*)0}$ final states. Continuum production of the $D_s^{(*)+}D_{sJ}^-$ signal in the $\Upsilon(2S)$ data sample was estimated by rescaling luminosities and correcting them for the c.m. energy dependence of the QED cross-section. These events were later eliminated from the $\Upsilon(2S)$ sample to acquire genuine QCD events. Interference between resonant and continuum amplitudes was not considered here \cite{PhysRevD.105.114001}. The statistical significance of $D_{s1}(2536)^{-}$ and $D_{s2}^*(2573)^{-}$ estimated with likelihood scan are respectively 6.8$\sigma$ and 4.0$\sigma$ in the resonant data sample; 18.3$\sigma$ and 10.1$\sigma$ in the continuum data sample.

    Observation of these signal events allows calculation of the branching fractions of $\Upsilon(2S)\to D_s^{(*)+}D_{sJ}^-$ as:    
    \begin{equation}
        \begin{split}
            \mathcal{B}(\Upsilon(2S) \rightarrow D_s^{(*)+}D_{s}^-)\mathcal{B}(D_{sJ}^- &   \rightarrow\bar{K}\bar{D}^{(*)}) = \\ & = \frac{N_{\Upsilon(2S)}^{\mathrm{sig}}}{N_{\Upsilon(2S)}\sum\varepsilon_i \mathcal{B}_i};
        \end{split}
        \label{eq:brfr}
    \end{equation} \\ and the Born cross-section of $e^+e^-\to D_s^{(*)+}D_{sJ}^-$ as: 
    \begin{equation}
        \begin{split}
            \sigma^{\mathrm{B}}(e^+e^-\rightarrow D_s^{(*)+} D_{sJ}^-) & \mathcal{B}(D_{sJ}^-\rightarrow\bar{K}\bar{D}^{(*)}) = \\ & = \frac{N_{\mathrm{cont}}^{\mathrm{sig}}|1-\Pi|^2}{\mathcal{L}_{\mathrm{cont}} (1+\delta_{\mathrm{ISR}}) \sum\varepsilon_i \mathcal{B}_i},
        \end{split}
        \label{eq:crosssect}
    \end{equation}
    where $i$ indicates the mode of $D_s$ decay, $\varepsilon_i$ is the respective reconstruction efficiencies, $\mathcal{L}_{cont}$ is the luminosity of the off-resonant sample, $|1-\Pi|^2$ is the vacuum polarization correction factor and $(1+\delta_{\mathrm{ISR}})$ is the ISR correction factor. The number of corrected signal events in the $\Upsilon(2S)$ data sample is 20$\pm$12$\pm$2 for the $D_{s}^{(*)+}D_{s2}^*(2573)^{-}$ decay, from which a statistical significance of only 1.6$\sigma$ has been estimated. By integrating a likelihood versus the number of $D_{s}^{(*)+}D_{s2}^*(2573)^{-}$ signal events, its upper limit is set at 90\% C.L. This calculation derives an upper limit on the production in $\Upsilon(2S)$ decays: $\mathcal{B}^{\mathrm{UL}}(\Upsilon(2S) \rightarrow D_s^{*+} D_{s 2}^*(2573)^-) \mathcal{B}(D_{s2}^*(2573)^- \rightarrow K^- \bar{D}^0)<2.9\cdot10^{-5}$. Similarly, the following upper limit has been set at 90\% C.L.: $\mathcal{B}^{\mathrm{UL}}(\Upsilon(2S) \rightarrow D_s^{*+} D_{s2}^*(2573)^{-}) \mathcal{B}(D_{s 2}^*(2573)^{-} \rightarrow K_S^0 D^{-})<3.0\cdot10^{-5}$. A listing of all numeric results for the branching fractions and the Born cross-sections calculations for the different decay modes can be found in the original source \cite{DsDsinY2S}. 

    Comparison of the $\Upsilon(2S)\to D_s^{(*)+} D_{sJ}^-$ process to the $e^+e^-/\Upsilon(2S)\to\mu^+\mu^-$ process indicates dominance of QCD-ruled production of the $D_s^{(*)+} D_{sJ}^-$ in the $\Upsilon(2S)$ decays. Calculation of the following fractions indicated good agreement with expected values of 0.498 and 0.497 from isospin symmetry:
    \begin{equation}
       \begin{split}
            \frac{\mathcal{B}(D_{s 1}(2536)^-\rightarrow K_S^0D^*(2010)^-)}{\mathcal{B}(D_{s 1}(2536)^{-} \rightarrow K^{-} D^*(2007)^0)} & = 0.48 \pm 0.07 \pm 0.02, \\
            \frac{\mathcal{B}(D_{s 2}^*(2573)^{-} \rightarrow K_S^0 D^{-})}{ \mathcal{B}(D_{s 2}^*(2573)^{-} \rightarrow K^{-} D^0)} & =0.49 \pm 0.10 \pm 0.02,
        \end{split} 
    \end{equation}

    \section{Study of $\mathbf{e^+e^-\to D_s^+D_{s0}^*(2317)^-A}$ + c.c. and $\mathbf{e^+e^- \to D_s^+ D_{s1}(2460)^-A}$ + c.c at Belle}

    A recent LHCb study, which reported observation of 7 new $X$ exotic states in $J/\psi\phi$ the system in the $B^-\to J/\psi\phi K^-$ process, heated-up the interest in search for exotics with hidden charm. In this view, an analysis of the inclusive $e^+e^-\to D_s^+D_{sJ}^{(*)-}A$ process in continuum has been proposed in Belle, where $A$ stands for “anything else” indicating inclusiveness of the process under study, and the following $D_{sJ}^{(*)-}$ mesons were considered: $D_{s0}^*(2317)^-$ and $D_{s1}(2460)^-$. The target of this study is to search for resonant states in the invariant mass systems of $D_s^+D_{s0}^*(2317)^-$ and $D_s^+D_{s1}(2460)^-$ in the continuum with the whole Belle data sets and to evaluate the cross-sections of the inclusive $e^+e^- \rightarrow X A$ processes, where $X$ are the states observed by LHCb, which satisfy quantum numbers' requirements.

    The reconstruction of primary and secondary $D_s^+$ mesons included $\phi K^+$, $\phi\rho^+$ and $K^*(982)K^+$ decay modes for both processes under study. Search for the $D_{s0}^*(2317)^-$ candidates has been performed in $D_s^-\pi^0$ combinations, while the $D_s^{*-}$ candidates were searched for in the $D_s^-\gamma$ combinations and further combined with $\pi^0$ candidates to acquire $D_{s1}(2460)^-$. A designated cut-based selection has been developed to validate the analysis procedure, but was later abandoned in favour of MVA-based approach to maximize selection efficiency. Further results have shown that MVA approach helped to increase the $D_{sJ}^{(*)-}$ yield by a factor of two. 

\begin{figure*}[!hbt]
        \begin{center}
            \includegraphics[width=0.8\textwidth]{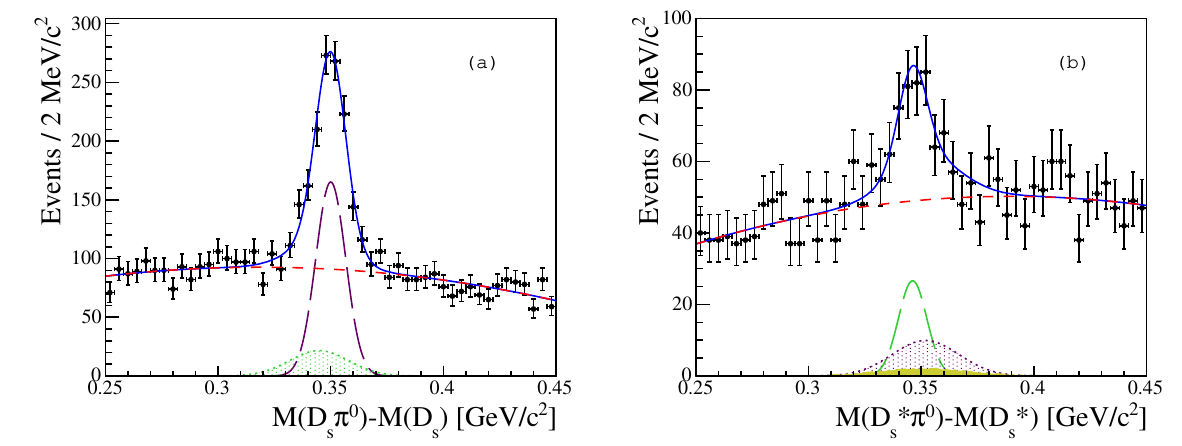}
            \caption{Simultaneous fit results on data sample, using the MLP approach. Figure (a) shows the $\Delta M(D_s\pi^0)$ distribution. The purple longdashed line represents here the true $D_{s0}^*(2317)$ signal contribution, while the green dotted area represents reflection of the $D_{s1}(2460)$. Figure (b) shows the $\Delta M(D_{s1}(2460))$ distribution. The green longdashed line here represents the true $D_{s1}(2460)$ signal contribution, the purple dotted area represents $D_{s0}^*(2317)$ reflection, and the solid-filled yellow area indicates broken signal contribution. The blue continuous line in both figures represents the result of the final fit. The red dashed line shows the polynomial fit to the generic background.}
            \label{fig:simultaneousCMC}
        \end{center}
    \end{figure*}

    Current analysis is strongly affected by a complicated background picture. The $e^+e^-\to D_s^+D_{s0}^*(2317)^-A$ and $e^+e^- \to D_s^+ D_{s1}(2460)^-A$ processes are qualitatively and kinematical similar, which leads to existence of the cross-feed background. Thus, multiple $D_{s0}^*(2317)^-$ candidates can be mistakenly combined of the $D_{s1}(2460)^-$ decay products, and vice versa. This leads to the existence of peaking background contribution in $\Delta M$ distributions of both. In addition, a so-called broken signal peaking background can be observed in the $\Delta M(D_{s1}(2460)^-)$ distribution because of possible combination of the $D_{s1}(2460)^-$ decay products, where a photon radiated by $D_s^{*-}$ is chosen incorrectly. The aforementioned sources of peaking background have been precisely studied in signal and generic MC. Parameters of the corresponding peaks were determined there and fixed for the simultaneous fit on data. Fig.~\ref{fig:simultaneousCMC} shows results of this fit. 

    \begin{figure*}[!hbt]
        \begin{center}
        \includegraphics[width=0.37\textwidth]{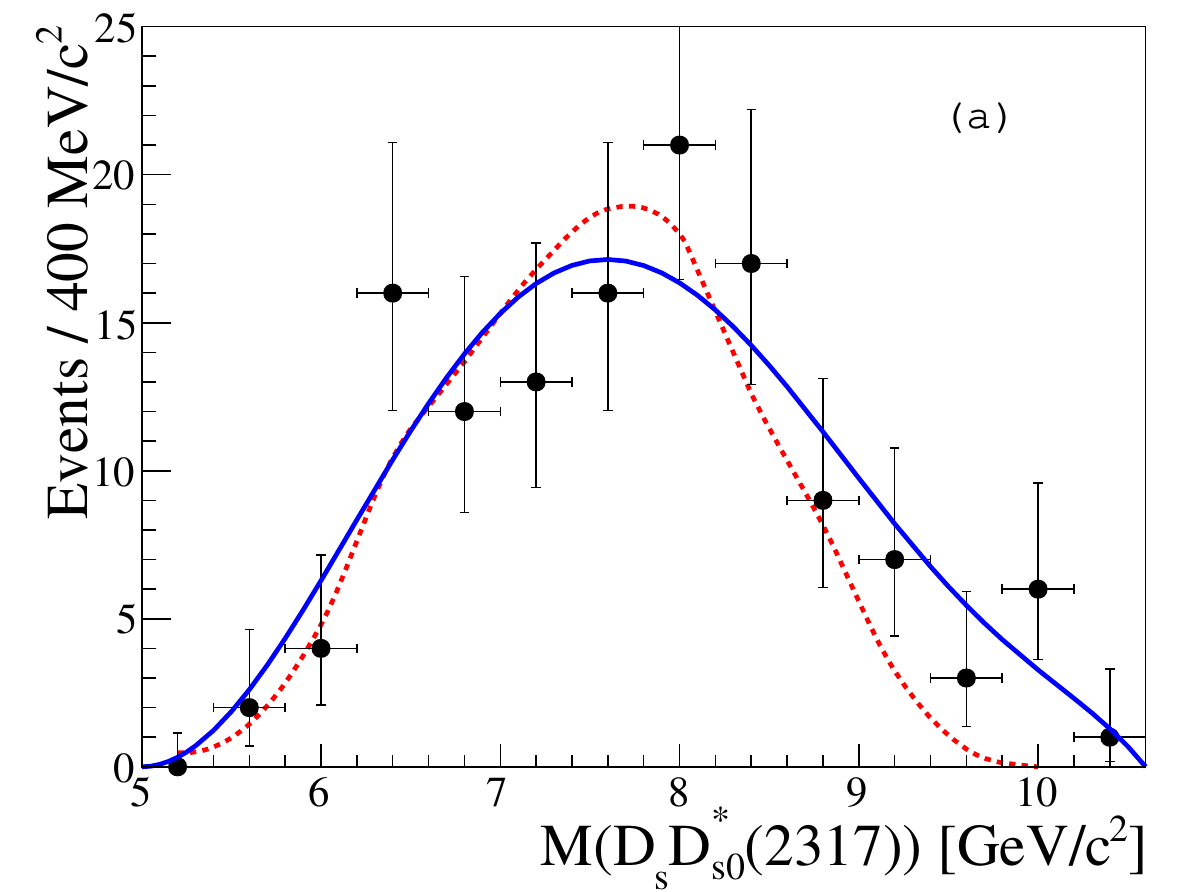}
        \includegraphics[width=0.37\textwidth]{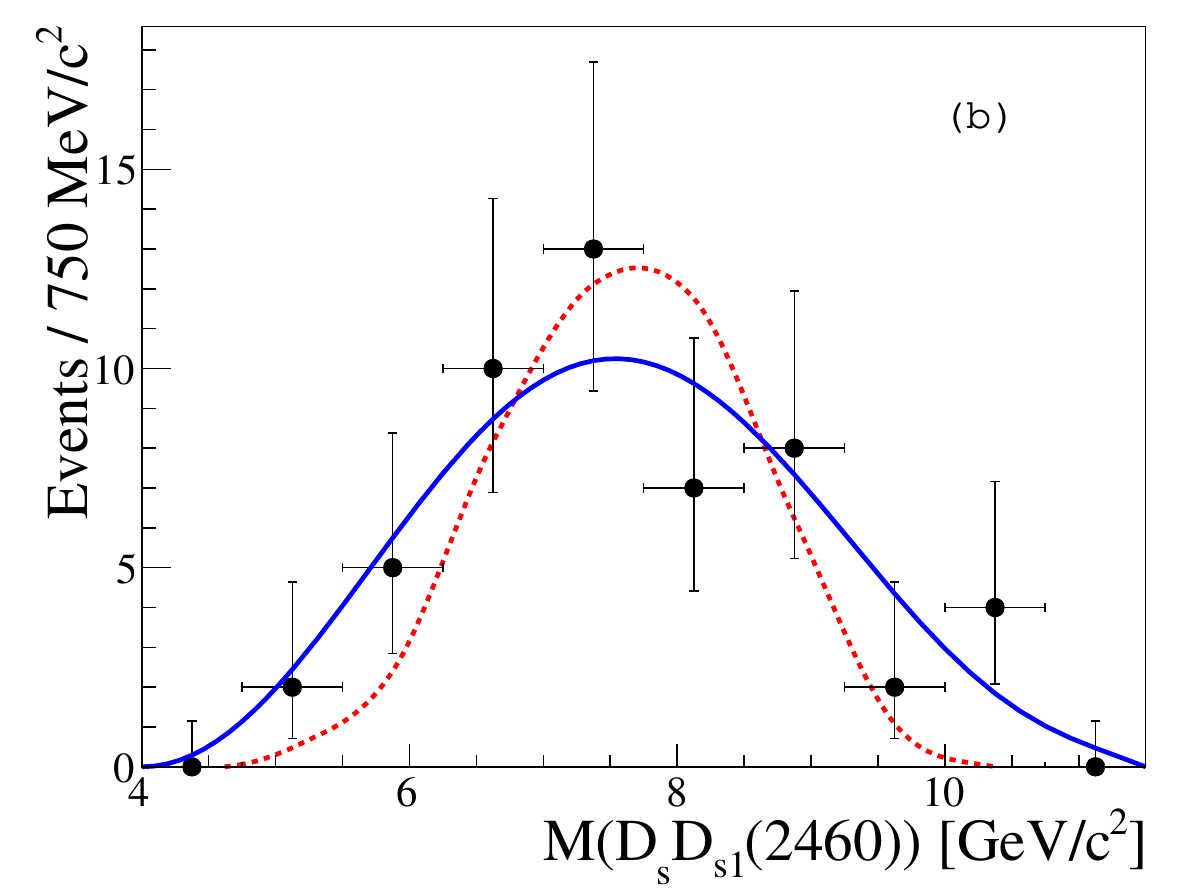}
        \caption{(a) $D_sD_{s0}^*(2317)$ and (b) $D_sD_{s1}(2460)$ invariant mass distributions on generic MC (red dotted line) and data (black point). The fit to data is performed using a Bernstein polynomial of $8^{th}$ order and is depicted with a blue continuous line.}
        \label{fig:DsDs2317}
        \end{center}
    \end{figure*}

    The measurement of the mass splitting is crucial for a better understanding of the nature of the $D_{s0}^{*-}(2317)$  and the $D_{s1}^{-}(2460)$ resonant states. These states can be interpreted as the first chiral partners of $c\bar{s}$ hadrons. As such, they represent rather a pattern of spontaneous breakdown of chiral symmetry than isolated events \cite{chilral}. The spontaneous chiral symmetry breaking yields a mass splitting between the chiral doublet of about 345 MeV/$c^2$ when the pion coupling to the doublet is half its coupling to a free quark. We found that the mass splitting of the $D_{s0}^{*-}(2317)$ meson is quantified in (350.0 $\pm$ 0.5) MeV/c$^2$. For comparison, the PDG2023 reports:  (349.4 $\pm$ 0.5) MeV/c$^2$. For the $D_{s1}^{-}(2460)$ meson, we found: (346.2 $\pm$ 1.7) MeV/c$^2$. For comparison, PDG2023 reports: (347.3 $\pm$ 0.7) MeV/c$^2$. Both results are in reasonable agreement with the former studies. The measured $D_{s0}^*(2317)^-$ and the $D_{s1}(2460)^-$ mass splitting resolution is (6.64 $\pm$ 0.53) MeV/$c^2$ and (6.27 $\pm$ 1.55) MeV/$c^2$, respectively. For comparison, a former Belle analysis in the continuum, where only one $cs$ meson was involved, delivered (7.1$\pm$0.2 MeV/$c^2$) and (7.6$\pm$0.5 MeV/$c^2$), respectively.

    Using the acquired $D_{s0}^{*-}(2317)$ and $D_{s1}(2460)^-$ yield, the following calculation has been performed:
    
    \begin{equation}
        \begin{split}
            \dfrac{Br(D_{s1}(2460)\to D_s^*\pi^0)}{Br(D_{s0}^*(2317)\to D_s\pi^0)} \times  \dfrac{\sigma (D_{s1}(2460), \text{MVA})}{\sigma (D_{s0}^*(2317), \text{MVA})} & = \\ = 0.26 \pm 0.07 \text{(stat)} \pm 0.03 \text{(syst)}&
        \end{split}
    \end{equation}

    This result is in agreement with the former Belle publication on  $e^+e^- \rightarrow D_{s0}^*(2317)^- A$ \cite{mikami}, though according to the quark model it should be a factor 10 higher.

    Fig.~\ref{fig:DsDs2317} shows on MC samples and data the invariant mass distributions of  $D_s^- D_{s0}^*(2317)^+$ and $D_s^- D_{s1}(2460)^+$. We observe higher fluctuations in data than on the MC samples due to the limited statistics. Still, we can conclude that the MC distributions and the data distributions match well, and there is no significant signal for a resonant state in the mass range under examination up to an integrated luminosity of 980.15 fb$^{-1}$.

    \begin{table}[!hbt]
        \begin{center}
            \begin{tabular}{c|c|c|c}
                \hline
                \hline
                \multirow{2}{*}{Decay chain} & Total & Estimated & $\sigma^{UL}\times$
                \\
                & error (\%) & $N^{UL}_{90}$ & $\mathcal{B}(X\to D_s D_{sJ}^*)$\\
                \hline
     
                 $e^+e^- \to X(4274)A$ & 13.3 & 2.45 & 122.5 \\
                 $e^+e^- \to X(4685)A$ & 14.1 & 2.04 & 101.8 \\
                 $e^+e^- \to X(4630)A$ & 18.3 & 2.05 & 228.1 \\
                 $e^+e^- \to X(4500)A$ & 18.0 & 2.34 & 260.1 \\
                 $e^+e^- \to X(4700)A$ & 18.7 & 2.18 & 241.8 \\
                \hline
                \hline
            \end{tabular}
            \caption{Results of $N^{UL}$ at 90\% C.L. and $\sigma^{UL}\times\mathcal{B}(X\to D_sD_{sJ}^{(*)})$ calculation for the considered processes. Total uncertainties are included in the calculation. QCD corrections, quantum polarization terms, and ISR contributions are not included as explained in the text. We report the visible cross-section UL at 90\% C.L., according to the formula in Eq. \ref{eq:crosssect}.}
            \label{tab:XstatesUL2}
        \end{center}
    \end{table}

    A custom tool based on the counting model \cite{ref40} had to be developed for the UL calculations to include systematic uncertainties correctly. With systematic uncertainties set to zero, the results were consistent with those acquired with a standard Feldman-Cousin tool \cite{macroroot}. The final results of this study are summarized in Tab.~\ref{tab:XstatesUL2}. Born cross-section ULs were evaluated using the equation, similar to Eq. \ref{eq:crosssect} with the only substitution of $\Sigma_i\mathcal{B}_i\varepsilon_i$ into $\Sigma_{ij}\mathcal{B}_i\mathcal{B}_i\varepsilon_{ij}$, where $\mathcal{B}_i$ and $\mathcal{B}_j$ denote branching fractions of primary and secondary $D_s$ mesons, and $\varepsilon_{ij}$ denotes the respective reconstruction efficiency. The results of the cross-section ULs estimated for the accessible $X$ states are summarized in Tab.~\ref{tab:XstatesUL2}.
    
    \section{Summary and Outlook}

    %This short overview features three analysis, which rely on Belle data to search for exotic hidden-charm candidates in continuum. The results reported by the authors affirm the importance of larger data samples available for studies of rare decays. This could allow reaching higher significance for the observed resonances and look for even more suppressed processes. It is claimed by the authors of featured studies that it is feasible to revisit these analyses, when a larger Belle II data sample is available.

    Results from three analyses to search for exotic hidden-charm candidates in the $e^+e^-$ continuum using Belle data were reported. Compared to $B$ meson decays, different quantum numbers of exotic states are accessible. By exploiting this advantage, various decay modes were studied by the authors, intending to measure respective cross-sections (UL) or/and to search for exotic candidates. Even though no statistically significant observation of any unconventional states is reported in this overview, side results offer scientific significance.
    
    In conclusion, authors affirm the importance of larger data samples available for studies of rare decays. Presently, data taking with the Belle II experiment is ongoing and is planned to provide a factor $\leq$50 larger integrated luminosity. This could allow reaching higher significance for the observed resonances and look for even more suppressed processes. It is claimed by the authors of featured studies that it is feasible to revisit these analyses, when a larger Belle II data sample is available.
    
    \bibliographystyle{apsrev4-1}
    \bibliography{main}

\begin{thebibliography}{37}%
\makeatletter
\providecommand \@ifxundefined [1]{%
 \@ifx{#1\undefined}
}%
\providecommand \@ifnum [1]{%
 \ifnum #1\expandafter \@firstoftwo
 \else \expandafter \@secondoftwo
 \fi
}%
\providecommand \@ifx [1]{%
 \ifx #1\expandafter \@firstoftwo
 \else \expandafter \@secondoftwo
 \fi
}%
\providecommand \natexlab [1]{#1}%
\providecommand \enquote  [1]{``#1''}%
\providecommand \bibnamefont  [1]{#1}%
\providecommand \bibfnamefont [1]{#1}%
\providecommand \citenamefont [1]{#1}%
\providecommand \href@noop [0]{\@secondoftwo}%
\providecommand \href [0]{\begingroup \@sanitize@url \@href}%
\providecommand \@href[1]{\@@startlink{#1}\@@href}%
\providecommand \@@href[1]{\endgroup#1\@@endlink}%
\providecommand \@sanitize@url [0]{\catcode `\\12\catcode `\$12\catcode `\&12\catcode `\#12\catcode `\^12\catcode `\_12\catcode `\%12\relax}%
\providecommand \@@startlink[1]{}%
\providecommand \@@endlink[0]{}%
\providecommand \url  [0]{\begingroup\@sanitize@url \@url }%
\providecommand \@url [1]{\endgroup\@href {#1}{\urlprefix }}%
\providecommand \urlprefix  [0]{URL }%
\providecommand \Eprint [0]{\href }%
\providecommand \doibase [0]{http://dx.doi.org/}%
\providecommand \selectlanguage [0]{\@gobble}%
\providecommand \bibinfo  [0]{\@secondoftwo}%
\providecommand \bibfield  [0]{\@secondoftwo}%
\providecommand \translation [1]{[#1]}%
\providecommand \BibitemOpen [0]{}%
\providecommand \bibitemStop [0]{}%
\providecommand \bibitemNoStop [0]{.\EOS\space}%
\providecommand \EOS [0]{\spacefactor3000\relax}%
\providecommand \BibitemShut  [1]{\csname bibitem#1\endcsname}%
\let\auto@bib@innerbib\@empty
%</preamble>
\bibitem [{\citenamefont {Brambilla}\ \emph {et~al.}(2020)\citenamefont {Brambilla} \emph {et~al.}}]{Brambilla_2020}%
  \BibitemOpen
  \bibfield  {author} {\bibinfo {author} {\bibfnamefont {N.}~\bibnamefont {Brambilla}} \emph {et~al.},\ }\href {\doibase https://doi.org/10.1016/j.physrep.2020.05.001} {\bibfield  {journal} {\bibinfo  {journal} {Physics Reports}\ }\textbf {\bibinfo {volume} {873}},\ \bibinfo {pages} {1} (\bibinfo {year} {2020})},\ \bibinfo {note} {the XYZ states: experimental and theoretical status and perspectives}\BibitemShut {NoStop}%
\bibitem [{\citenamefont {Aubert}\ \emph {et~al.}(2005)\citenamefont {Aubert} \emph {et~al.}}]{Y4260_BaBar}%
  \BibitemOpen
  \bibfield  {author} {\bibinfo {author} {\bibfnamefont {B.}~\bibnamefont {Aubert}} \emph {et~al.} (\bibinfo {collaboration} {BABAR Collaboration}),\ }\href {\doibase 10.1103/PhysRevLett.95.142001} {\bibfield  {journal} {\bibinfo  {journal} {Phys. Rev. Lett.}\ }\textbf {\bibinfo {volume} {95}},\ \bibinfo {pages} {142001} (\bibinfo {year} {2005})}\BibitemShut {NoStop}%
\bibitem [{\citenamefont {Aubert}\ \emph {et~al.}(2007)\citenamefont {Aubert} \emph {et~al.}}]{Y4360_BaBar}%
  \BibitemOpen
  \bibfield  {author} {\bibinfo {author} {\bibfnamefont {B.}~\bibnamefont {Aubert}} \emph {et~al.} (\bibinfo {collaboration} {BABAR Collaboration}),\ }\href {\doibase 10.1103/PhysRevLett.98.212001} {\bibfield  {journal} {\bibinfo  {journal} {Phys. Rev. Lett.}\ }\textbf {\bibinfo {volume} {98}},\ \bibinfo {pages} {212001} (\bibinfo {year} {2007})}\BibitemShut {NoStop}%
\bibitem [{\citenamefont {Yuan}\ \emph {et~al.}(2007{\natexlab{a}})\citenamefont {Yuan} \emph {et~al.}}]{Y4360_Belle}%
  \BibitemOpen
  \bibfield  {author} {\bibinfo {author} {\bibfnamefont {C.~Z.}\ \bibnamefont {Yuan}} \emph {et~al.} (\bibinfo {collaboration} {Belle Collaboration}),\ }\href {\doibase 10.1103/PhysRevLett.99.142002} {\bibfield  {journal} {\bibinfo  {journal} {Phys. Rev. Lett.}\ }\textbf {\bibinfo {volume} {99}},\ \bibinfo {pages} {142002} (\bibinfo {year} {2007}{\natexlab{a}})}\BibitemShut {NoStop}%
\bibitem [{\citenamefont {He}\ \emph {et~al.}(2006)\citenamefont {He} \emph {et~al.}}]{Y4260_CLEO}%
  \BibitemOpen
  \bibfield  {author} {\bibinfo {author} {\bibfnamefont {Q.}~\bibnamefont {He}} \emph {et~al.} (\bibinfo {collaboration} {CLEO Collaboration}),\ }\href {\doibase 10.1103/PhysRevD.74.091104} {\bibfield  {journal} {\bibinfo  {journal} {Phys. Rev. D}\ }\textbf {\bibinfo {volume} {74}},\ \bibinfo {pages} {091104} (\bibinfo {year} {2006})}\BibitemShut {NoStop}%
\bibitem [{\citenamefont {Yuan}\ \emph {et~al.}(2007{\natexlab{b}})\citenamefont {Yuan} \emph {et~al.}}]{Y4260_Belle}%
  \BibitemOpen
  \bibfield  {author} {\bibinfo {author} {\bibfnamefont {C.~Z.}\ \bibnamefont {Yuan}} \emph {et~al.} (\bibinfo {collaboration} {Belle Collaboration}),\ }\href {\doibase 10.1103/PhysRevLett.99.182004} {\bibfield  {journal} {\bibinfo  {journal} {Phys. Rev. Lett.}\ }\textbf {\bibinfo {volume} {99}},\ \bibinfo {pages} {182004} (\bibinfo {year} {2007}{\natexlab{b}})}\BibitemShut {NoStop}%
\bibitem [{\citenamefont {Chiu}\ \emph {et~al.}(2006)\citenamefont {Chiu} \emph {et~al.}}]{latticeQCD_Y}%
  \BibitemOpen
  \bibfield  {author} {\bibinfo {author} {\bibfnamefont {T.-W.}\ \bibnamefont {Chiu}} \emph {et~al.} (\bibinfo {collaboration} {TWQCD Collaboration}),\ }\href {\doibase 10.1103/PhysRevD.73.111503} {\bibfield  {journal} {\bibinfo  {journal} {Physical Review D}\ }\textbf {\bibinfo {volume} {73}},\ \bibinfo {pages} {4} (\bibinfo {year} {2006})}\BibitemShut {NoStop}%
\bibitem [{\citenamefont {Ablikim}\ \emph {et~al.}(2017{\natexlab{a}})\citenamefont {Ablikim} \emph {et~al.}}]{Y4230_BSIII}%
  \BibitemOpen
  \bibfield  {author} {\bibinfo {author} {\bibfnamefont {M.}~\bibnamefont {Ablikim}} \emph {et~al.} (\bibinfo {collaboration} {BESIII Collaboration}),\ }\href {\doibase 10.1103/PhysRevLett.118.092001} {\bibfield  {journal} {\bibinfo  {journal} {Phys. Rev. Lett.}\ }\textbf {\bibinfo {volume} {118}},\ \bibinfo {pages} {092001} (\bibinfo {year} {2017}{\natexlab{a}})}\BibitemShut {NoStop}%
\bibitem [{\citenamefont {Ablikim}\ \emph {et~al.}(2017{\natexlab{b}})\citenamefont {Ablikim} \emph {et~al.}}]{Y4230_BESIII_confirm_1}%
  \BibitemOpen
  \bibfield  {author} {\bibinfo {author} {\bibfnamefont {M.}~\bibnamefont {Ablikim}} \emph {et~al.} (\bibinfo {collaboration} {BESIII Collaboration}),\ }\href {\doibase 10.1103/PhysRevLett.118.092002} {\bibfield  {journal} {\bibinfo  {journal} {Phys. Rev. Lett.}\ }\textbf {\bibinfo {volume} {118}},\ \bibinfo {pages} {092002} (\bibinfo {year} {2017}{\natexlab{b}})}\BibitemShut {NoStop}%
\bibitem [{\citenamefont {Ablikim}\ \emph {et~al.}(2014)\citenamefont {Ablikim} \emph {et~al.}}]{Y4230_BESIII_confirm_2}%
  \BibitemOpen
  \bibfield  {author} {\bibinfo {author} {\bibfnamefont {M.}~\bibnamefont {Ablikim}} \emph {et~al.},\ }\href {\doibase 10.1088/1674-1137/38/4/043001} {\bibfield  {journal} {\bibinfo  {journal} {Chinese Physics C}\ }\textbf {\bibinfo {volume} {38}},\ \bibinfo {pages} {043001} (\bibinfo {year} {2014})}\BibitemShut {NoStop}%
\bibitem [{\citenamefont {Ablikim}\ \emph {et~al.}(2019{\natexlab{a}})\citenamefont {Ablikim} \emph {et~al.}}]{Y4230_BESIII_chicomega}%
  \BibitemOpen
  \bibfield  {author} {\bibinfo {author} {\bibfnamefont {M.}~\bibnamefont {Ablikim}} \emph {et~al.} (\bibinfo {collaboration} {BESIII Collaboration}),\ }\href {\doibase 10.1103/PhysRevD.99.091103} {\bibfield  {journal} {\bibinfo  {journal} {Phys. Rev. D}\ }\textbf {\bibinfo {volume} {99}},\ \bibinfo {pages} {091103} (\bibinfo {year} {2019}{\natexlab{a}})}\BibitemShut {NoStop}%
\bibitem [{\citenamefont {Ablikim}\ \emph {et~al.}(2019{\natexlab{b}})\citenamefont {Ablikim} \emph {et~al.}}]{Y4230_BESIII_pibarDDst}%
  \BibitemOpen
  \bibfield  {author} {\bibinfo {author} {\bibfnamefont {M.}~\bibnamefont {Ablikim}} \emph {et~al.} (\bibinfo {collaboration} {BESIII Collaboration}),\ }\href {\doibase 10.1103/PhysRevLett.122.102002} {\bibfield  {journal} {\bibinfo  {journal} {Phys. Rev. Lett.}\ }\textbf {\bibinfo {volume} {122}},\ \bibinfo {pages} {102002} (\bibinfo {year} {2019}{\natexlab{b}})}\BibitemShut {NoStop}%
\bibitem [{\citenamefont {Ablikim}\ \emph {et~al.}(2020)\citenamefont {Ablikim} \emph {et~al.}}]{Y4230_BESIII_etaJpsi}%
  \BibitemOpen
  \bibfield  {author} {\bibinfo {author} {\bibfnamefont {M.}~\bibnamefont {Ablikim}} \emph {et~al.} (\bibinfo {collaboration} {BESIII Collaboration}),\ }\href {\doibase 10.1103/PhysRevD.102.031101} {\bibfield  {journal} {\bibinfo  {journal} {Phys. Rev. D}\ }\textbf {\bibinfo {volume} {102}},\ \bibinfo {pages} {031101} (\bibinfo {year} {2020})}\BibitemShut {NoStop}%
\bibitem [{\citenamefont {Zhu}(2022)}]{Y4230_BESIII_etaprimJpsi}%
  \BibitemOpen
  \bibfield  {author} {\bibinfo {author} {\bibfnamefont {K.}~\bibnamefont {Zhu}},\ }\href {\doibase 10.1103/PhysRevD.105.L031506} {\bibfield  {journal} {\bibinfo  {journal} {Phys. Rev. D}\ }\textbf {\bibinfo {volume} {105}},\ \bibinfo {pages} {L031506} (\bibinfo {year} {2022})}\BibitemShut {NoStop}%
\bibitem [{\citenamefont {Ablikim}\ \emph {et~al.}(2018)\citenamefont {Ablikim} \emph {et~al.}}]{BESIII_KKJpsi}%
  \BibitemOpen
  \bibfield  {author} {\bibinfo {author} {\bibfnamefont {M.}~\bibnamefont {Ablikim}} \emph {et~al.} (\bibinfo {collaboration} {BESIII Collaboration}),\ }\href {\doibase 10.1103/PhysRevD.97.071101} {\bibfield  {journal} {\bibinfo  {journal} {Phys. Rev. D}\ }\textbf {\bibinfo {volume} {97}},\ \bibinfo {pages} {071101} (\bibinfo {year} {2018})}\BibitemShut {NoStop}%
\bibitem [{\citenamefont {Jia}\ \emph {et~al.}(2019)\citenamefont {Jia} \emph {et~al.}}]{SenJia_1}%
  \BibitemOpen
  \bibfield  {author} {\bibinfo {author} {\bibfnamefont {S.}~\bibnamefont {Jia}} \emph {et~al.} (\bibinfo {collaboration} {Belle Collaboration}),\ }\href {\doibase 10.1103/PhysRevD.100.111103} {\bibfield  {journal} {\bibinfo  {journal} {Phys. Rev. D}\ }\textbf {\bibinfo {volume} {100}},\ \bibinfo {pages} {111103} (\bibinfo {year} {2019})}\BibitemShut {NoStop}%
\bibitem [{\citenamefont {Jia}\ \emph {et~al.}(2020)\citenamefont {Jia} \emph {et~al.}}]{SenJia_2}%
  \BibitemOpen
  \bibfield  {author} {\bibinfo {author} {\bibfnamefont {S.}~\bibnamefont {Jia}} \emph {et~al.} (\bibinfo {collaboration} {Belle Collaboration}),\ }\href {\doibase 10.1103/PhysRevD.101.091101} {\bibfield  {journal} {\bibinfo  {journal} {Phys. Rev. D}\ }\textbf {\bibinfo {volume} {101}},\ \bibinfo {pages} {091101} (\bibinfo {year} {2020})}\BibitemShut {NoStop}%
\bibitem [{\citenamefont {Aaij}\ \emph {et~al.}(2017)\citenamefont {Aaij} \emph {et~al.}}]{LHCb_jpsiPhiK_1}%
  \BibitemOpen
  \bibfield  {author} {\bibinfo {author} {\bibfnamefont {R.}~\bibnamefont {Aaij}} \emph {et~al.} (\bibinfo {collaboration} {LHCb Collaboration}),\ }\href {\doibase 10.1103/PhysRevLett.118.022003} {\bibfield  {journal} {\bibinfo  {journal} {Phys. Rev. Lett.}\ }\textbf {\bibinfo {volume} {118}},\ \bibinfo {pages} {022003} (\bibinfo {year} {2017})}\BibitemShut {NoStop}%
\bibitem [{\citenamefont {Aaij}\ \emph {et~al.}(2021)\citenamefont {Aaij} \emph {et~al.}}]{LHCb_jpsiPhiK_2}%
  \BibitemOpen
  \bibfield  {author} {\bibinfo {author} {\bibfnamefont {R.}~\bibnamefont {Aaij}} \emph {et~al.} (\bibinfo {collaboration} {LHCb Collaboration}),\ }\href {\doibase 10.1103/PhysRevLett.127.082001} {\bibfield  {journal} {\bibinfo  {journal} {Phys. Rev. Lett.}\ }\textbf {\bibinfo {volume} {127}},\ \bibinfo {pages} {082001} (\bibinfo {year} {2021})}\BibitemShut {NoStop}%
\bibitem [{CMS(2020)}]{CMS_X6900}%
  \BibitemOpen
  \href {https://pos.sissa.it/390/} {\emph {\bibinfo {title} {{Proceedings, 40th International Conference on High Energy Physics (ICHEP2020)}}}}\ (\bibinfo  {publisher} {SISSA},\ \bibinfo {year} {2020})\BibitemShut {NoStop}%
\bibitem [{\citenamefont {Chen}\ \emph {et~al.}(2022)\citenamefont {Chen} \emph {et~al.}}]{doubl_charmonium_1}%
  \BibitemOpen
  \bibfield  {author} {\bibinfo {author} {\bibfnamefont {H.-X.}\ \bibnamefont {Chen}} \emph {et~al.},\ }\href {\doibase 10.1088/1361-6633/aca3b6} {\bibfield  {journal} {\bibinfo  {journal} {Reports on Progress in Physics}\ }\textbf {\bibinfo {volume} {86}},\ \bibinfo {pages} {026201} (\bibinfo {year} {2022})}\BibitemShut {NoStop}%
\bibitem [{\citenamefont {Sang}\ \emph {et~al.}(2023)\citenamefont {Sang} \emph {et~al.}}]{doubl_charmonium_2}%
  \BibitemOpen
  \bibfield  {author} {\bibinfo {author} {\bibfnamefont {W.-L.}\ \bibnamefont {Sang}} \emph {et~al.},\ }\href@noop {} {\enquote {\bibinfo {title} {Electromagnetic and hadronic decay of fully heavy tetraquark},}\ } (\bibinfo {year} {2023}),\ \Eprint {http://arxiv.org/abs/2307.16150} {arXiv:2307.16150 [hep-ph]} \BibitemShut {NoStop}%
\bibitem [{\citenamefont {Aubert}\ \emph {et~al.}(2010)\citenamefont {Aubert} \emph {et~al.}}]{BaBar_DstX_crosssect}%
  \BibitemOpen
  \bibfield  {author} {\bibinfo {author} {\bibfnamefont {B.}~\bibnamefont {Aubert}} \emph {et~al.} (\bibinfo {collaboration} {The BABAR Collaboration}),\ }\href {\doibase 10.1103/PhysRevD.81.011102} {\bibfield  {journal} {\bibinfo  {journal} {Phys. Rev. D}\ }\textbf {\bibinfo {volume} {81}},\ \bibinfo {pages} {011102} (\bibinfo {year} {2010})}\BibitemShut {NoStop}%
\bibitem [{\citenamefont {Zhang}\ and\ \citenamefont {Chao}(2008)}]{DstX_theoryRef}%
  \BibitemOpen
  \bibfield  {author} {\bibinfo {author} {\bibfnamefont {Y.-J.}\ \bibnamefont {Zhang}}\ and\ \bibinfo {author} {\bibfnamefont {K.-T.}\ \bibnamefont {Chao}},\ }\href {\doibase 10.1103/PhysRevD.78.094017} {\bibfield  {journal} {\bibinfo  {journal} {Phys. Rev. D}\ }\textbf {\bibinfo {volume} {78}},\ \bibinfo {pages} {094017} (\bibinfo {year} {2008})}\BibitemShut {NoStop}%
\bibitem [{\citenamefont {Karliner}\ \emph {et~al.}(2016)\citenamefont {Karliner} \emph {et~al.}}]{eta_exchange}%
  \BibitemOpen
  \bibfield  {author} {\bibinfo {author} {\bibfnamefont {M.}~\bibnamefont {Karliner}} \emph {et~al.},\ }\href {\doibase https://doi.org/10.1016/j.nuclphysa.2016.03.057} {\bibfield  {journal} {\bibinfo  {journal} {Nuclear Physics A}\ }\textbf {\bibinfo {volume} {954}},\ \bibinfo {pages} {365} (\bibinfo {year} {2016})}\BibitemShut {NoStop}%
\bibitem [{\citenamefont {Yin}\ \emph {et~al.}(2023)\citenamefont {Yin} \emph {et~al.}}]{etacJpsi}%
  \BibitemOpen
  \bibfield  {author} {\bibinfo {author} {\bibfnamefont {J.~H.}\ \bibnamefont {Yin}} \emph {et~al.},\ }\href@noop {} {\enquote {\bibinfo {title} {Search for the double-charmonium state with $\eta_c {J}/\psi$ at {B}elle},}\ } (\bibinfo {year} {2023}),\ \Eprint {http://arxiv.org/abs/2305.17947} {arXiv:2305.17947 [hep-ex]} \BibitemShut {NoStop}%
\bibitem [{\citenamefont {Abashian}\ \emph {et~al.}(2002)\citenamefont {Abashian} \emph {et~al.}}]{Belle_detctor}%
  \BibitemOpen
  \bibfield  {author} {\bibinfo {author} {\bibfnamefont {A.}~\bibnamefont {Abashian}} \emph {et~al.},\ }\href {\doibase https://doi.org/10.1016/S0168-9002(01)02013-7} {\bibfield  {journal} {\bibinfo  {journal} {Nuclear Instruments and Methods in Physics Research Section A: Accelerators, Spectrometers, Detectors and Associated Equipment}\ }\textbf {\bibinfo {volume} {479}},\ \bibinfo {pages} {117} (\bibinfo {year} {2002})},\ \bibinfo {note} {detectors for Asymmetric B-factories}\BibitemShut {NoStop}%
\bibitem [{\citenamefont {Zhang}\ \emph {et~al.}(2023)\citenamefont {Zhang} \emph {et~al.}}]{zhang2023twoloop}%
  \BibitemOpen
  \bibfield  {author} {\bibinfo {author} {\bibfnamefont {Y.-D.}\ \bibnamefont {Zhang}} \emph {et~al.},\ }\href@noop {} {\enquote {\bibinfo {title} {Two-loop {QCD} corrections to c even bottomonium exclusive decays to double ${J}/\psi$},}\ } (\bibinfo {year} {2023}),\ \Eprint {http://arxiv.org/abs/2310.07453} {arXiv:2310.07453 [hep-ph]} \BibitemShut {NoStop}%
\bibitem [{\citenamefont {Abe}\ \emph {et~al.}(2007)\citenamefont {Abe} \emph {et~al.}}]{PhysRevLett.98.082001}%
  \BibitemOpen
  \bibfield  {author} {\bibinfo {author} {\bibfnamefont {K.}~\bibnamefont {Abe}} \emph {et~al.} (\bibinfo {collaboration} {Belle Collaboration}),\ }\href {\doibase 10.1103/PhysRevLett.98.082001} {\bibfield  {journal} {\bibinfo  {journal} {Phys. Rev. Lett.}\ }\textbf {\bibinfo {volume} {98}},\ \bibinfo {pages} {082001} (\bibinfo {year} {2007})}\BibitemShut {NoStop}%
\bibitem [{\citenamefont {Yang}\ \emph {et~al.}(2014)\citenamefont {Yang} \emph {et~al.}}]{PhysRevD.90.112008}%
  \BibitemOpen
  \bibfield  {author} {\bibinfo {author} {\bibfnamefont {S.~D.}\ \bibnamefont {Yang}} \emph {et~al.} (\bibinfo {collaboration} {Belle Collaboration}),\ }\href {\doibase 10.1103/PhysRevD.90.112008} {\bibfield  {journal} {\bibinfo  {journal} {Phys. Rev. D}\ }\textbf {\bibinfo {volume} {90}},\ \bibinfo {pages} {112008} (\bibinfo {year} {2014})}\BibitemShut {NoStop}%
\bibitem [{\citenamefont {Benayoun}\ \emph {et~al.}(1999)\citenamefont {Benayoun} \emph {et~al.}}]{BENAYOUN_1999}%
  \BibitemOpen
  \bibfield  {author} {\bibinfo {author} {\bibfnamefont {M.}~\bibnamefont {Benayoun}} \emph {et~al.},\ }\href {\doibase 10.1142/s021773239900273x} {\bibfield  {journal} {\bibinfo  {journal} {Modern Physics Letters A}\ }\textbf {\bibinfo {volume} {14}},\ \bibinfo {pages} {2605–2614} (\bibinfo {year} {1999})}\BibitemShut {NoStop}%
\bibitem [{\citenamefont {Gao}\ \emph {et~al.}(2023)\citenamefont {Gao} \emph {et~al.}}]{DsDsinY2S}%
  \BibitemOpen
  \bibfield  {author} {\bibinfo {author} {\bibfnamefont {B.~S.}\ \bibnamefont {Gao}} \emph {et~al.} (\bibinfo {collaboration} {Belle Collaboration}),\ }\href {\doibase 10.1103/PhysRevD.108.112015} {\bibfield  {journal} {\bibinfo  {journal} {Phys. Rev. D}\ }\textbf {\bibinfo {volume} {108}},\ \bibinfo {pages} {112015} (\bibinfo {year} {2023})}\BibitemShut {NoStop}%
\bibitem [{\citenamefont {Guo}\ \emph {et~al.}(2022)\citenamefont {Guo} \emph {et~al.}}]{PhysRevD.105.114001}%
  \BibitemOpen
  \bibfield  {author} {\bibinfo {author} {\bibfnamefont {Y.~P.}\ \bibnamefont {Guo}} \emph {et~al.},\ }\href {\doibase 10.1103/PhysRevD.105.114001} {\bibfield  {journal} {\bibinfo  {journal} {Phys. Rev. D}\ }\textbf {\bibinfo {volume} {105}},\ \bibinfo {pages} {114001} (\bibinfo {year} {2022})}\BibitemShut {NoStop}%
\bibitem [{\citenamefont {Nowak}\ \emph {et~al.}(1993)\citenamefont {Nowak} \emph {et~al.}}]{chilral}%
  \BibitemOpen
  \bibfield  {author} {\bibinfo {author} {\bibfnamefont {M.~A.}\ \bibnamefont {Nowak}} \emph {et~al.},\ }\href {\doibase 10.1103/PhysRevD.48.4370} {\bibfield  {journal} {\bibinfo  {journal} {Phys. Rev. D}\ }\textbf {\bibinfo {volume} {48}},\ \bibinfo {pages} {4370} (\bibinfo {year} {1993})}\BibitemShut {NoStop}%
\bibitem [{\citenamefont {Mikami}\ \emph {et~al.}(2004)\citenamefont {Mikami} \emph {et~al.}}]{mikami}%
  \BibitemOpen
  \bibfield  {author} {\bibinfo {author} {\bibfnamefont {Y.}~\bibnamefont {Mikami}} \emph {et~al.} (\bibinfo {collaboration} {Belle Collaboration}),\ }\href {\doibase 10.1103/PhysRevLett.92.012002} {\bibfield  {journal} {\bibinfo  {journal} {Phys. Rev. Lett.}\ }\textbf {\bibinfo {volume} {92}},\ \bibinfo {pages} {012002} (\bibinfo {year} {2004})}\BibitemShut {NoStop}%
\bibitem [{\citenamefont {Read}(2002)}]{ref40}%
  \BibitemOpen
  \bibfield  {author} {\bibinfo {author} {\bibfnamefont {A.~L.}\ \bibnamefont {Read}},\ }\href {\doibase 10.1088/0954-3899/28/10/313} {\bibfield  {journal} {\bibinfo  {journal} {Journal of Physics G: Nuclear and Particle Physics}\ }\textbf {\bibinfo {volume} {28}},\ \bibinfo {pages} {2693} (\bibinfo {year} {2002})}\BibitemShut {NoStop}%
\bibitem [{mac()}]{macroroot}%
  \BibitemOpen
  \href@noop {} {\enquote {\bibinfo {title} {Root class tfeldmancousin.c},}\ }\bibinfo {howpublished} {\url{https://root.cern/doc/master/}},\ \bibinfo {note} {accessed: 2023-12-01}\BibitemShut {NoStop}%
\end{thebibliography}%
\end{document}